\documentclass{sf2a-conf2017}
\usepackage{graphicx}
\usepackage{xcolor}
\usepackage[colorlinks = true, citecolor = red]{hyperref}
\usepackage[]{natbib}  
\usepackage{epstopdf}

\def\BibTeX{{\rm B\kern-.05em{\sc i\kern-.025em b}\kern-.08em
    T\kern-.1667em\lower.7ex\hbox{E}\kern-.125emX}}
\bibpunct{(}{)}{;}{a}{}{,}  

\begin{document}

\TitreGlobal{proceeding of the SF2A 2017. 7 pages, 2 Fig}

\title{New insights on the origin of the High Velocity Peaks in the Galactic Bulge}

\runningtitle{Milky Way bulge}

\author{J. G. Fern\'andez-Trincado}\address{Departamento de Astronom\'\i a, Casilla 160-C, Universidad de Concepci\'on, Concepci\'on, Chile \\ email: {\tt \url{jfernandezt@astro-udec.cl} and/or \url{jfernandezt87@gmail.com}}}

\author{A. C. Robin}\address{Institut Utinam, CNRS UMR 6213, Universit\'e Bourgogne-Franche-Comt\'e, OSU THETA Franche-Comt\'e, Observatoire de Besan\c{c}on, BP 1615, 25010 Besan\c{c}on Cedex, France}

\author{E. Moreno}\address{Instituto de Astronom\'ia, Universidad Nacional Aut\'onoma de M\'exico, Apdo. Postal 70264, M\'exico D.F., 04510, M\'exico}

\author{A. P\'erez-Villegas}\address{Max-Planck-Institut f\"ur Extraterrestrische Physik, Gie\ss enbachstra\ss e, 85748 Garching, Germany}

\author{B. Pichardo$^{3}$}


\setcounter{page}{237}

\maketitle


\begin{abstract}
We provide new insight on the origin of the cold high-V$_{\rm los}$ peaks ($\sim$200 kms$^{-1}$) in the Milky Way bulge discovered in the APOGEE commissioning data \citep{Nidever2012}. Here we show that such kinematic behaviour present in the field regions towards the Galactic bulge is not likely associated with orbits that build the boxy/peanut (B/P) bulge. To this purpose, a new set of test particle simulations of a kinematically cold stellar disk evolved in a 3D steady-state barred Milky Way galactic potential, has been analysed in detail. Especially bar particles trapped into the bar are identified through the orbital Jacobi energy $E_{J}$, which allows us to identify the building blocks of the B/P feature and investigate their kinematic properties. Finally, we present preliminary results showing that the high-V$_{\rm los}$ features observed towards the Milky Way bulge are a natural consequence of a large-scale \textit{midplane} particle structure, which is unlikely associated with the Galactic bar. 
\end{abstract}


\section{Introduction}

The discovery by \citet[][]{Nidever2012} of cold high-velocity peaks ($\sim$ 200 km s$^{-1}$) in the Apache Point Observatory Galactic Evolution Experiment (APOGEE) commissioning data across the Galactic bulge $l=$ \{4, 14\} and $b=$ \{-2, 2\} and confirmed by the High-Order Kinematic Moments by \citet{Zasowski2016}, suggests that there may be a significant non-axisymmetric structure that dominates the bulge regions \citep[e.g.,][for instance]{Robin2012, Wegg2013}, which has turned the study and characterization of a B/P bulge \citep[e.g.,][among others]{Portail2015, Simion2017} into a very active research field. Most of models that attempt to explain the high-velocity peaks observed toward the Milky Way bulge suggest that these features are most likely bulge stars on bar orbits, i.e., orbits in 2:1 and/or higher order resonant family \citep[see e.g.,][]{Aumer2015, Molloy2015}. Recently, alternative scenarios have been proposed that do not invoke any family of bar resonant orbits linked with the building blocks of the B/P feature. Additionally, it has been suggested that the high-velocity peaks may be the product of a kiloparsec-scale nuclear stellar disk in the Galactic bulge \citep[][]{Debattista2015}. Also, the recent study by \citet[][]{Li2014} suggests that these kinematics features might be an artifact due to small number statistics. With these issues in mind, we expect this preliminary contribution will help improve the current understanding on the origin of the cold high-velocity peaks. In this work, we qualitatively analyzed a set of numerical simulations of a synthetic Milky Way Galaxy made up of the superposition of many composite stellar populations already described and analyzed in \citet{Fernandez-Trincado2017Thesis}. Using numerical simulations from \citet{Fernandez-Trincado2017Thesis}, we began a pilot project aimed to provide an alternative scenario for the origin of  high-V$_{\rm los}$ feature in the bulge to look for possible orbital energy imprints of the cold high-V$_{\rm los}$. 
 
 \begin{figure}[ht!]
 	\centering
 	\includegraphics[width=0.60\textwidth,clip]{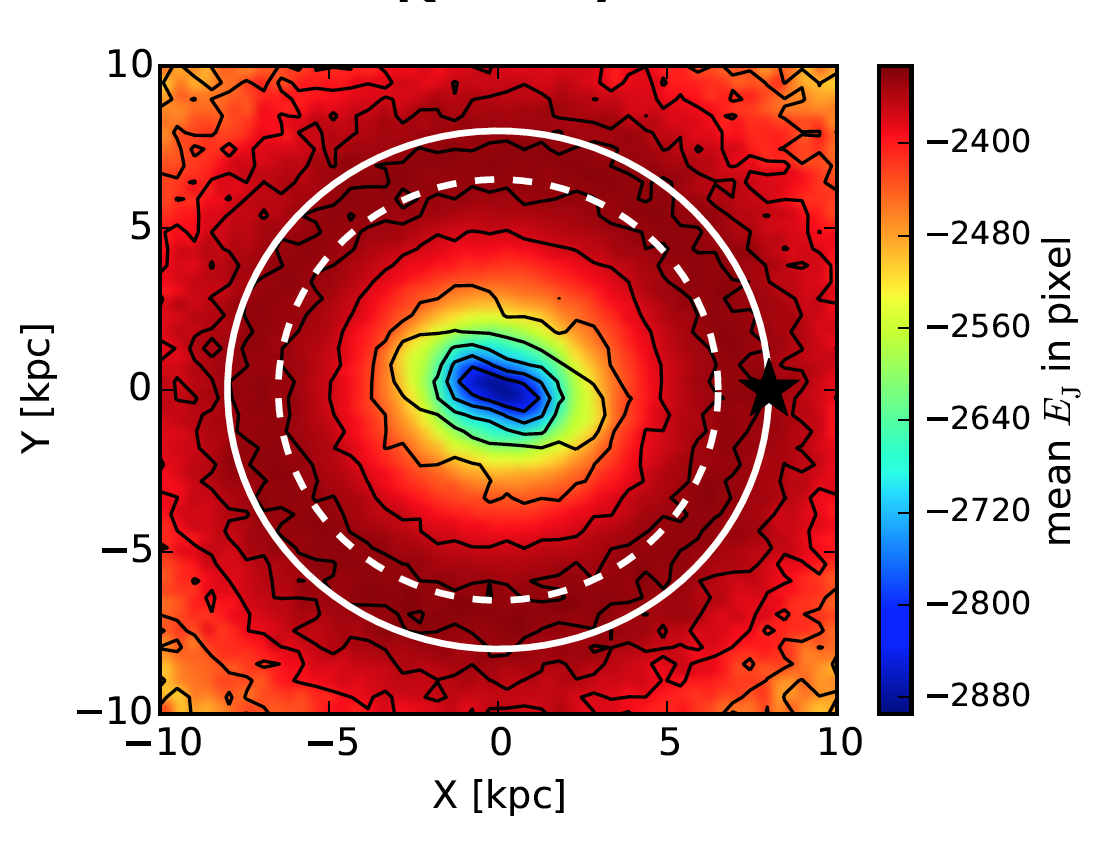}%
 	\caption{Face-on view of the simulated cold stellar thin disk in the \textit{inertial} reference frame where the bar is at an angle of 20 degrees from the Sun-GC line of sight. Colors indicate $\langle E_{J} \rangle$ in units of 100 km$^{2}$ s$^{-2}$. The white dashed circle indicates the corotation radius (6.5 kpc), which is in good agreement with results from the literature to explain the Hercules group \citet{Perez-Villegas2017}; and the white circle marks the solar radius (8 kpc) and the present-day solar position (black star symbol). The black contours refer to the surface density distribution for the entire sample of 1$\times$10$^{6}$ particles in t$=$15 Gyr. }
 	\label{figure1}
 \end{figure}
 
\section{The Galactic Model} 
 
 We use the galactic dynamic software \textit{GravPot16}\footnote{\url{https://fernandez-trincado.github.io/GravPot16/}} in order to carry out a comprehensive orbital study of particles in the inner region of the Milky Way. For a more detailed discussion about \textit{GravPot16}, we refer the readers to a forthcoming paper (Fern\'andez-Trincado et al. in preparation). Here we summarize the backbone of \textit{GravPot16}. Using the new version of the Besan\c{c}on Galaxy Model, in good agreement with many observations, we computed a semi-analytical steady-state 3D gravitational potential of the Milky Way, observationally and dynamically constrained. The model is primarily made up of the superposition of several composite stellar components, where the density profiles in cylindrical coordinates, $\rho_i$(R, Z), are the same as those proposed in \citet[][]{Robin2003, Robin2012, Robin2014}, i.e., a B/P bulge, a Hernquist stellar halo, seven stellar Einasto thin disks with spherical symmetry in the inner regions, two stellar sech$^{2}$ thick disks, a gaseous exponential disk, and a spherical structure associated with the dark matter halo. A new formulation for the global potential, $\Phi$(R, Z),  of this Milky Way density model, $\Sigma\rho_i$(R, Z), will be described in detail in a forthcoming paper (Fern\'andez-Trincado, et al. in preparation). $\Phi$(R, Z) has been rescaled to the Sun's galactocentric distance. The Sun is located at $R_{\odot}=$ 8.0 kpc, and the local rotation velocity is assumed to be $\Theta_0({\rm R_{\odot}}) = 244.5$ km s$^{-1}$, given by \citet{Sofue2015}. Here, we briefly describe the bar's structural parameters, such as recommended by \citet{Fernandez-Trincado2017Thesis} from dynamical constraints using the BRAVA data set \citep{Kunder2012}: We assume a total mass for the bar of 1.1$\times$10$^{10}$ M$_{\odot}$, an angle of 20 degrees for the present-day orientation of the major axis of the bar and an angular velocity, $\Omega_{bar} = 35$ km s$^{-1}$ kpc$^{-1}$, consistent with the recent estimate of \citet[][]{Portail2015}, and a cut-off radius $R_{c}=3.28$ kpc \citep[e.g.,][]{Robin2012}. Additionally, it should be noted that the non-axisymmetric configuration of our dynamic model has been extensively employed to predict stellar orbits \citep[see][]{Fernandez-Trincado2016b, Fernandez-Trincado2017}, and/or orbital parameters for a large set of APOGEE-TGAS sources \citep[see][]{Abolfathi2017, Tang2017}. For a more detailed discussion, we refer the readers to a forthcoming paper \citep{Anders2017a, Anders2017b}. 

 \begin{figure}[ht!]
 	\centering
 	\includegraphics[width=1.0\textwidth,clip]{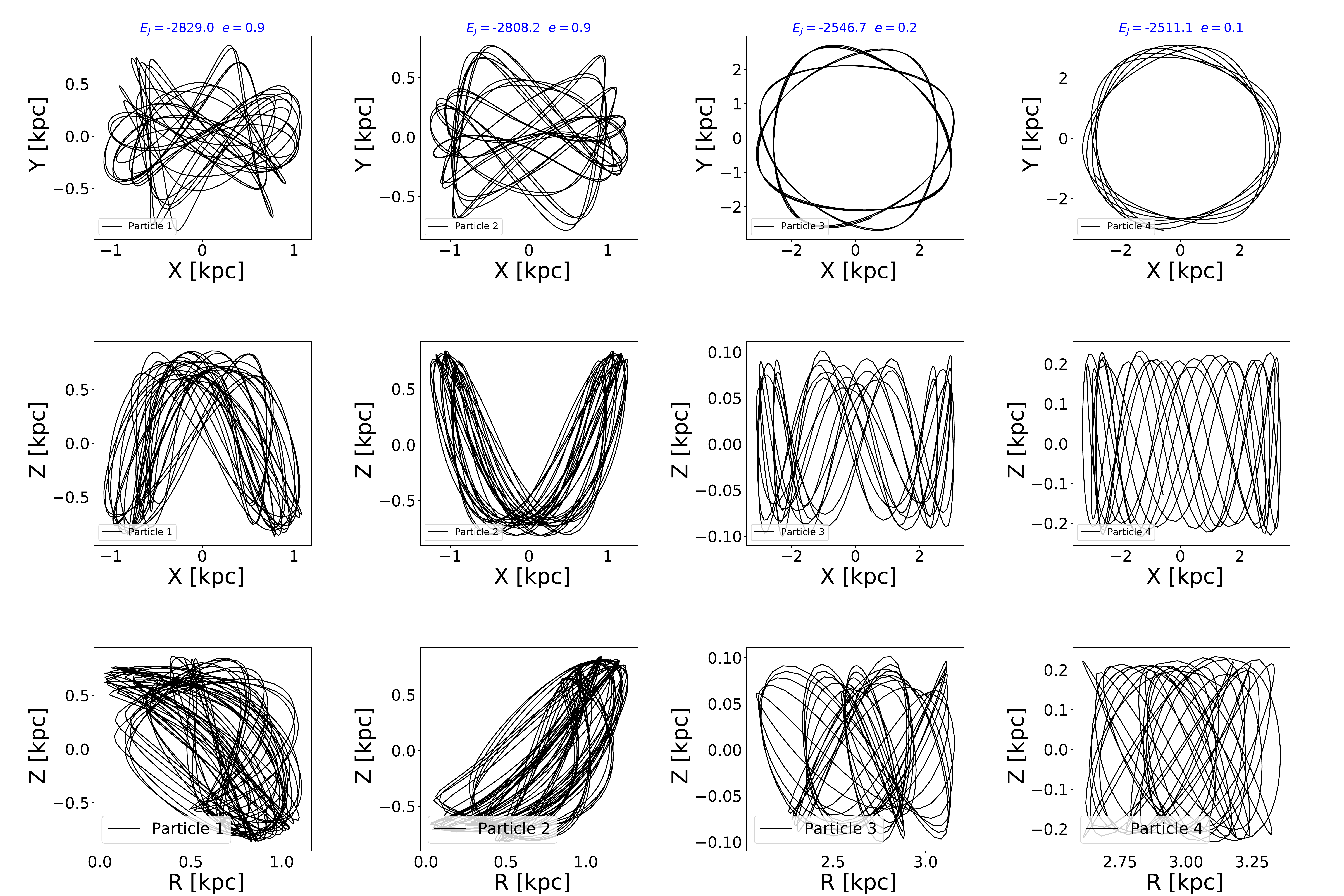}%
 	\caption{Orbits viewed face-on (top), side-on (middle) and meridional (bottom) in the \textit{non-inertial} reference frame where the bar is at rest. Column 1 and column 2 show the typical orbital configuration found for bar-trapped particles ($E_{J} < E_{J}^{boundary}$), while the column 3 and 4 show the orbital configuration for not-bar-trapped particles ($E_{J} > E_{J}^{boundary}$). The top blue label indicates the orbital eccentricity and its respective Jacobi energy ($E_{J}$).}
 	\label{figure2}
 \end{figure}

\section{Test particle simulations} 
 
First, we ran controled particle simulations to mimic one of the cold stellar thin disk described in the Besan\c{}con population synthesis model \citep[disk in the age range 7 to 10 Gyr; see e.g.,][]{Robin2003}. To this purpose we use the \textit{GravPot16} code in its axisymmetric and non-axisymmetric configuration. We adopt a similar strategy as described in \citet{Romero-Gomez2015} and \citet{LA2016}. The test particles are initially involved in a steady axisymmetric potential model over long integration time (in this work we adopt a integration time of 10 Gyr) to ensure that the initial disk particle distribution reaches a state of relaxation within the background potential. Then, the boxy bar structure grows adiabatically into the simulations during a period of time of 2 Gyr. Once the bar potential is introduced into the system, we increase the integration time during a period of time long enough ($> 3$ Gyr) to avoid transient effects. The initial conditions for the particle velocities are assigned using the Besan\c{c}on population synthesis model disc kinematics fitted to RAVE and TGAS data \citep[see e.g.,][]{Robin2017}. It is important to note that our initial conditions are based on locally self-consistent recipes, but it is not guaranteed to be fully self-consistent globally, and will thus be slightly relaxed before turning on the non-axisymmetric potential \citep[e.g.,][]{Fernandez-Trincado2017Thesis}. 

Secondly, after 15 Gyr integration time in the above potential we record the Jacobi energy per unit mass, $E_{J}$\footnote{The Jacobi energy is then given by $E_{J} =  \frac{1}{2}\vec{v}^2  + \Phi_{axi}(R, Z) + \Phi_{bar}(R, Z)  - \frac{1}{2}| \vec{\Omega}_{bar}\times \vec{R} |^2$. Where $\Phi_{axi}(R, Z)$ and $\Phi_{bar}(R, Z)$ are the \textit{GravPot16} axisymmetric and non-axisymmetric potential components, respectively.}, in the bar frame for all particles in a box of $\pm$3.5 kpc $\times$ 2.5 kpc $\times$ 2 kpc, which is thought to have high chance to contain orbits trapped into the bar structure as illustrated in Fig. \ref{figure1} and Fig. \ref{figure2} (first and second column). With the Jacobi energy distribution in the box above mentioned we determined the boundary between bar-trapped particles ($E_{J}< E_{J}^{boundary}$) and not-bar-trapped particles ($E_{J} > E_{J}^{boundary}$) by identifying the trough in the Jacobi energy distribution ($E_{J}^{boundary}\sim$ -2.7$\times$10$^{5}$ km$^{2}$ s$^{-2}$). The results are briefly described in \S\ref{results}.

 \begin{figure}[ht!]
 	\centering
 	\includegraphics[width=0.5\textwidth,clip]{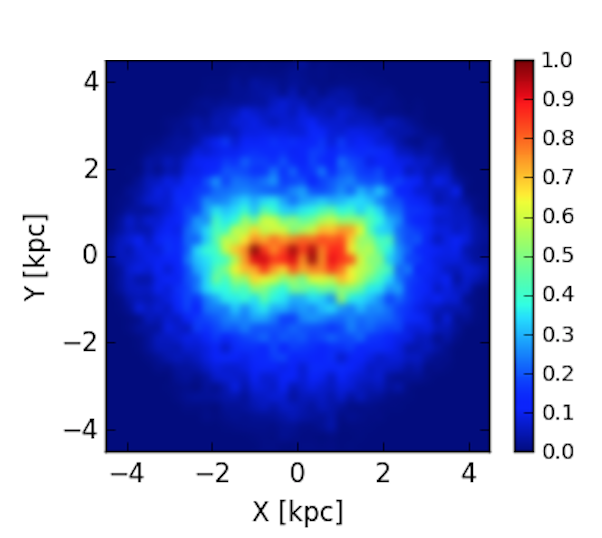}\includegraphics[width=0.47\textwidth,clip]{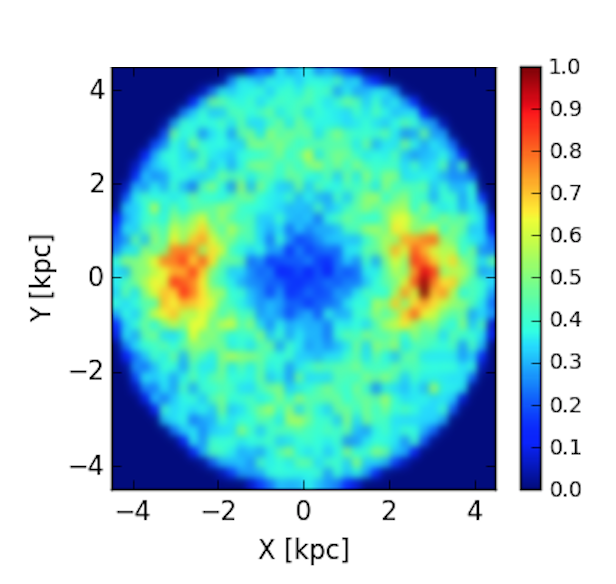}
 	\includegraphics[width=0.5\textwidth,clip]{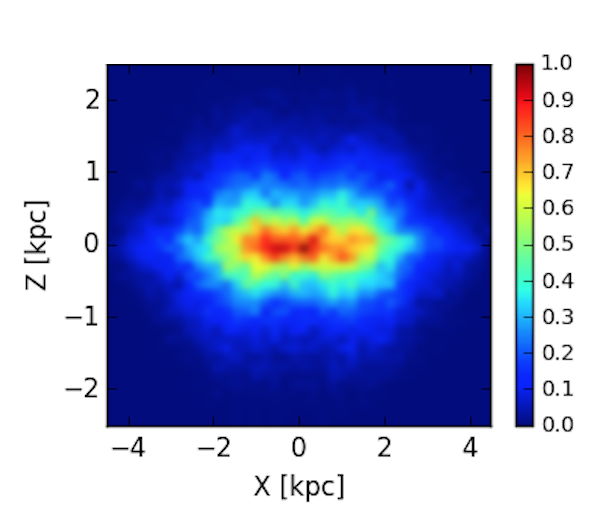}\includegraphics[width=0.47\textwidth,clip]{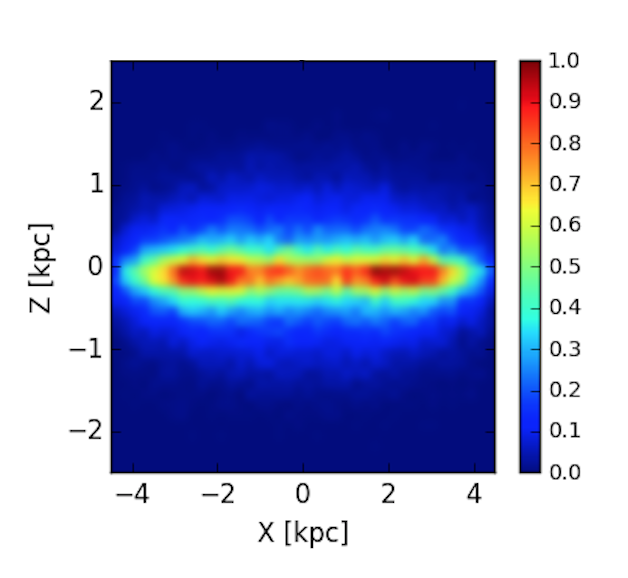}
 	\caption{ Kernel Density Estimate (KDE) smoothed distributions of bar-trapped particles (left column) and not-bar-trapped particles (right column) of the simulated cold stellar thin disk in the \textit{non-inertial} reference frame where the bar is at rest. Density distribution for the entire sample viewed face-on (first row) and side-on (second row). }
 	\label{figure3}
 \end{figure}

\section{Results and Concluding Remarks} 
 \label{results}
 
 Figure \ref{figure3} plots the  Kernel Density Estimate (KDE) smoothed distributions for the bar-trapped particles (first column). In particular we note that the B/P feature is carried largely by particles having a Jacobi energy $E_{J} < E_{J}^{boundary}$. In our numerical simulations, all the building blocks of the B/P bulge structure are composite of different orbits existing at energies smaller than the boundary energy, $E_{J} < E_{J}^{boundary}$, in particular diverse resonant orbits (i.e., family of tube orbits; $x_{1}v_{1}$: banana orbits, etc) which generate a strong peanut shape at shorter radii on the side-on projection (column 1, row 2 in the same figure). This Galactic B/P structure accounts for $\sim$34\% of the particles of the bulge (within $\sim 5$ kpc) \citep[e.g.,][]{Fernandez-Trincado2017Thesis}. Figure \ref{figure3} also plots the KDE smoothed distributions for the not-bar-trapped particles (second column), which do not show the B/P shape. The not-bar-trapped particles, those orbits existing at energies larger than the boundary energy, ($E_{J} > E_{J}^{boundary}$) in this model consist mostly of low eccentricity orbits, which dominate the mid-plane (see column 2 and 3 in Figure \ref{figure2}) and accounts for $\sim$66\% of the particles of the bulge. 

\subsection{An alternative explanation for the kinematics feature at high $V_{los}$} 

 It is important to note that here we provide the kinematic predictions for orbits existing at energies smaller or greater than  the boundary energy. Detailed azimuthal projections were already analyzed in \citet{Fernandez-Trincado2017Thesis} confirming the presence of the cold high-V$_{\rm los}$ peaks extending to Galactic longitude $l$ $\sim$ 10 degree, which are absent a few kiloparsecs off the mid-plane, indicating that orbits with Jacobi energy (a mid-plane hosting more particles at $E_{J} > E_{J}^{boundary}$) responsible for the feature do not extend this far off-plane, as also shown in second column in Figure \ref{figure3}. Figure \ref{figure4} plots the predicted line-of-sight velocity distributions (LOSVDs --here called V$_{\rm los}$) in the Galacto-centric restframe. At $E_{J} < E_{J}^{boundary}$ the V$_{\rm los}$ distribution has a single peak dominated by bar-trapped-particles, hosting more particles at $V_{los}<$ 150 km s$^{-1}$. There are also few particles that have high V$_{\rm los}$ ($\sim$ 200 km s$^{-1}$) and are likely associated with the high-velocity tail of the resonant bar-supporting 2:1 orbits \citep[see][]{Molloy2015}. At $E_{J} > E_{J}^{boundary}$ the V$_{\rm los}$ have developed two peaks, with particles moving at significantly larger velocities ($~200$ km s$^{-1}$) in the mid-plane, dominated by not-bar-trapped particles, but remain well below the circular velocity of the galaxy. These two high velocity peaks are more prominent than the low-V$_{\rm los}$ peak developed by bar-trapped particles.

\subsection{Conclusion} 

\begin{figure}[ht!]
	\centering
	\includegraphics[width=0.45\textwidth,clip]{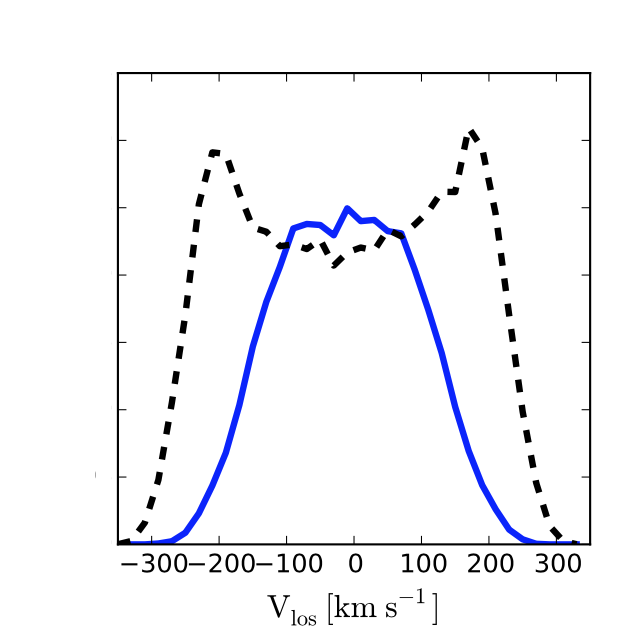}\includegraphics[width=0.57\textwidth,clip]{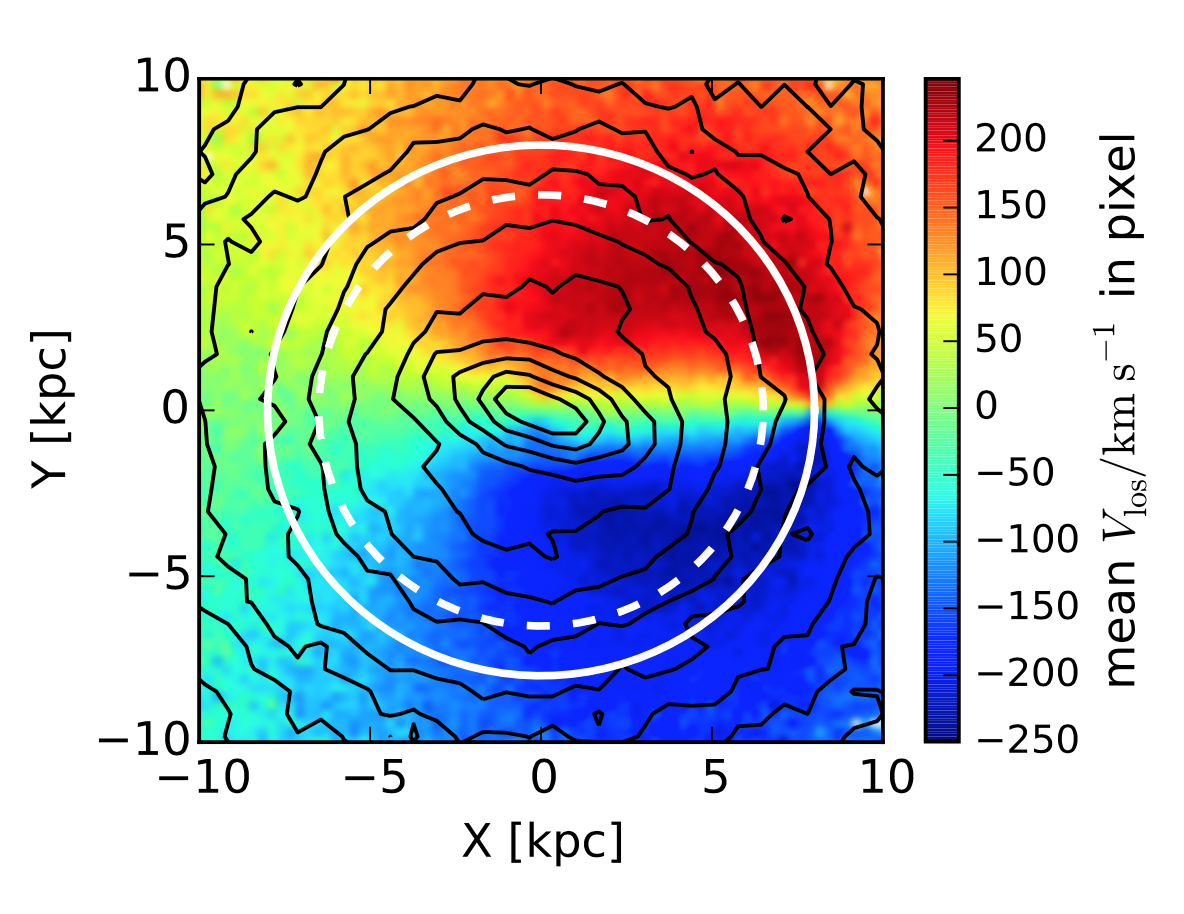}     
	\caption{\textit{Left}: $V_{los}$ histograms normalised to unit peak showing a dual-peak structure for the not-bar-trapped particles (black dashed line) and an unimodal distribution for the bar-trapped particles (blue line), using 20 km s$^{-1}$ binning. \textit{Right}: Kinematics map of the simulated cold stellar thin disk in Galactic coordinates for the entire sample of 1$\times$10$^{6}$ particles. The white circles and black contours levels are identical to those in Figure \ref{figure1}. Colors indicate $\langle V_{los} \rangle$.}
	\label{figure4}
\end{figure}

We have made an attempt to explain the presence of the cold high velocity peaks in the bulge. It is important to note, that we account for the composite nature of the bulge in our simulations. The dependence of V$_{\rm los}$ with $l$ and $b$ has not been shown in the present work, but extensively studied by \citet{Fernandez-Trincado2017Thesis}. The right panel of Figure \ref{figure4} shows color-coded maps of the average $V_{los}$, $\langle V_{los}\rangle$, for the entire sample of our simulated cold stellar thin disk in Galactic coordinates. In a similar manner as in \citet{Debattista2015} our numerical approach is capable of producing  the peak velocities at orbit tangent points with the characteristic \textit{winged} pattern of the velocity fields.

Lastly, we conclude that the most natural interpretation of the high velocity features towards the Galactic bulge is that they are likely not dominated by orbits at $E_{J} < E_{J}^{boundary}$ that build the B/P bulge, but may be a consequence of families of orbits at $E_{J} > E_{J}^{boundary}$ and low orbital eccentricities in the mid-plane that do not support the bar structure. Our Milky Way potential model fine-tuned to observations is able to explain the velocity distributions in most APOGEE fields in the bulge, without invoking the presence of any nuclear disk in the inner $\sim$ 1 kpc as pointed out in \citet{Debattista2015}. The advantage of our numerical approach is that the test particles have evolved in a realistic Milky Way potential inheriting the information on both density and kinematics, and the particles in statistical equilibrium with the potential imposed \citep[e.g.,][]{Romero-Gomez2015, LA2016, Fernandez-Trincado2017Thesis}. It should be noticed that we find very similar $V_{los}$ distributions to those in APOGEE, without any adjustment parameters, but without applying the observation selection function. Hence we shall verify this point in the near future.

The high precision of the \textit{Gaia} mission will provide the 6D phase space needed to confirm our orbital interpretations and to compute the orbital Jacobi energy beyond $\sim$ 5 kpc from the Sun.

\begin{acknowledgements}
	
We thank Merce Romero G\'omez, Fran\c{}coise Combes and Famaey Benoit for comments on an earlier version of this paper. J.G.F-T gratefully acknowledges partial financial support from the SF2A in order to attend the SF2A-2017 meeting held in Paris in July 2017, and the Chilean BASAL Centro de Excelencia en Astrof\'isica y Tecnolog\'ias Afines (CATA) grant PFB- 06/2007. We also acknowledge the support of the UTINAM Institute of the Universit\'e de Franche-Comt\'e, R\'egion de Franche-Comt\'e and Institut des Sciences de l'Univers (INSU) for providing HPC resources on the Cluster Supercomputer M\'esocentre de calcul de Franche-Comt\'e. E.M, A.P.V and B.P acknowledge support from UNAM/PAPIIT grant IN105916 and IN114114.
 
 Funding for the \textit{GravPot16} software has been provided by the Centre national d'\'etudes spatiale (CNES) through grant 0101973 and UTINAM Institute of the Universit\'e de Franche-Comte, supported by the Region de Franche-Comte and Institut des Sciences de l'Univers (INSU). The \textit{GravPot16} Web site is at \url{https://fernandez-trincado.github.io/GravPot16/}.

\end{acknowledgements}


\begin{thebibliography}{27}
	\expandafter\ifx\csname natexlab\endcsname\relax\def\natexlab#1{#1}\fi
	
	\bibitem[{{Abolfathi} {et~al.}(2017){Abolfathi}, {Aguado}, {Aguilar}, {Allende
			Prieto}, {Almeida}, {Tasnim Ananna}, {Anders}, {Anderson}, {Andrews},
		{Anguiano}, \& et~al.}]{Abolfathi2017}
	{Abolfathi}, B., {Aguado}, D.~S., {Aguilar}, G., {et~al.} 2017, ArXiv e-prints
	
	\bibitem[{{Anders} {et~al.}(2017{\natexlab{a}}){Anders}, {Queiroz},
		{Chiappini}, {Santiago}, {Fern\'andez-Trincado}, \& {Meza}}]{Anders2017b}
	{Anders}, F., {Queiroz}, A.~B., {Chiappini}, C., {et~al.} 2017{\natexlab{a}},
	in preparation
	
	\bibitem[{{Anders} {et~al.}(2017{\natexlab{b}}){Anders}, {Queiroz},
		{Chiappini}, {Santiago}, {Fern\'andez-Trincado}, \& {Meza}}]{Anders2017a}
	{Anders}, F., {Queiroz}, A.~B., {Chiappini}, C., {et~al.} 2017{\natexlab{b}},
	in IAU Symposium, Vol. 334, Rediscovering our Galaxy, ed. C.~{Chiappini},
	I.~{Minchev}, E.~{Starkenburg}, \& M.~{Valentini}
	
	\bibitem[{{Aumer} \& {Sch{\"o}nrich}(2015)}]{Aumer2015}
	{Aumer}, M. \& {Sch{\"o}nrich}, R. 2015, \mnras, 454, 3166
	
	\bibitem[{{Debattista} {et~al.}(2015){Debattista}, {Ness}, {Earp}, \&
		{Cole}}]{Debattista2015}
	{Debattista}, V.~P., {Ness}, M., {Earp}, S.~W.~F., \& {Cole}, D.~R. 2015,
	\apjl, 812, L16
	
		\bibitem[{{Fern{\'a}ndez-Trincado} {et~al.}(2017){Fern{\'a}ndez-Trincado},
			{Zamora}, {Garcia-Hernandez}, {Souto}, {Dell'Agli}, {Schiavon}, {Geisler},
			{Tang}, {Villanova}, {Hasselquist}, {Mennickent}, {Cunha}, {Shetrone},
			{Allende Prieto}, {Vieira}, {Zasowski}, {Sobeck}, {Hayes}, {Majewski},
			{Placco}, {Beers}, {Schleicher}, {Robin}, {Meszaros}, {Masseron}, {Garcia
				Perez}, {Anders}, {Meza}, {Alves-Brito}, {Carrera}, {Minniti}, {Lane},
			{Fernandez-Alvar}, {Moreno}, {Pichardo}, {Perez-Villegas}, {Schultheis},
			{Roman-Lopes}, {Fuentes}, {Nitschelm}, {Harding}, {Bizyaev}, {Pan},
			{Oravetz}, {Simmons}, {Ivans}, {Blanco-Cuaresma}, {Hernandez},
			{Alonso-Garcia}, {Valenzuela}, \& {Chaname}}]{Fernandez-Trincado2017}
		{Fern{\'a}ndez-Trincado}, J.~G., {Zamora}, O., {Garcia-Hernandez}, D.~A.,
		{et~al.} 2017b, ArXiv e-prints
		
	
	\bibitem[{{Fern\'andez-Trincado}(2017{\natexlab{a}})}]{Fernandez-Trincado2017Thesis}
	{Fern\'andez-Trincado}, J.~G. 2017{\natexlab{a}}, Universit\'e Bourgogne
	Franche-Comt\'e Phd Thesis, 1, 174
	

	\bibitem[{{Fern{\'a}ndez-Trincado} {et~al.}(2016){Fern{\'a}ndez-Trincado},
		{Robin}, {Moreno}, {Schiavon}, {Garc{\'{\i}}a P{\'e}rez}, {Vieira}, {Cunha},
		{Zamora}, {Sneden}, {Souto}, {Carrera}, {Johnson}, {Shetrone}, {Zasowski},
		{Garc{\'{\i}}a-Hern{\'a}ndez}, {Majewski}, {Reyl{\'e}}, {Blanco-Cuaresma},
		{Martinez-Medina}, {P{\'e}rez-Villegas}, {Valenzuela}, {Pichardo}, {Meza},
		{M{\'e}sz{\'a}ros}, {Sobeck}, {Geisler}, {Anders}, {Schultheis}, {Tang},
		{Roman-Lopes}, {Mennickent}, {Pan}, {Nitschelm}, \&
		{Allard}}]{Fernandez-Trincado2016b}
	{Fern{\'a}ndez-Trincado}, J.~G., {Robin}, A.~C., {Moreno}, E., {et~al.} 2016,
	\apj, 833, 132
	

	\bibitem[{{Kunder} {et~al.}(2012){Kunder}, {Koch}, {Rich}, {de Propris},
		{Howard}, {Stubbs}, {Johnson}, {Shen}, {Wang}, {Robin}, {Kormendy}, {Soto},
		{Frinchaboy}, {Reitzel}, {Zhao}, \& {Origlia}}]{Kunder2012}
	{Kunder}, A., {Koch}, A., {Rich}, R.~M., {et~al.} 2012, \aj, 143, 57
	
	\bibitem[{{Li} {et~al.}(2014){Li}, {Shen}, {Rich}, {Kunder}, \& {Mao}}]{Li2014}
	{Li}, Z.-Y., {Shen}, J., {Rich}, R.~M., {Kunder}, A., \& {Mao}, S. 2014, \apjl,
	785, L17
	
	\bibitem[{{Martinez-Medina} {et~al.}(2016){Martinez-Medina}, {Pichardo},
		{Moreno}, {Peimbert}, \& {Velazquez}}]{LA2016}
	{Martinez-Medina}, L.~A., {Pichardo}, B., {Moreno}, E., {Peimbert}, A., \&
	{Velazquez}, H. 2016, \apjl, 817, L3
	
	\bibitem[{{Molloy} {et~al.}(2015){Molloy}, {Smith}, {Evans}, \&
		{Shen}}]{Molloy2015}
	{Molloy}, M., {Smith}, M.~C., {Evans}, N.~W., \& {Shen}, J. 2015, \apj, 812,
	146
	
	\bibitem[{{Nidever} {et~al.}(2012){Nidever}, {Zasowski}, {Majewski}, {Bird},
		{Robin}, {Martinez-Valpuesta}, {Beaton}, {Sch{\"o}nrich}, {Schultheis},
		{Wilson}, {Skrutskie}, {O'Connell}, {Shetrone}, {Schiavon}, {Johnson},
		{Weiner}, {Gerhard}, {Schneider}, {Allende Prieto}, {Sellgren}, {Bizyaev},
		{Brewington}, {Brinkmann}, {Eisenstein}, {Frinchaboy}, {Garc{\'{\i}}a
			P{\'e}rez}, {Holtzman}, {Hearty}, {Malanushenko}, {Malanushenko}, {Muna},
		{Oravetz}, {Pan}, {Simmons}, {Snedden}, \& {Weaver}}]{Nidever2012}
	{Nidever}, D.~L., {Zasowski}, G., {Majewski}, S.~R., {et~al.} 2012, \apjl, 755,
	L25
	
	\bibitem[{{P{\'e}rez-Villegas} {et~al.}(2017){P{\'e}rez-Villegas}, {Portail},
		{Wegg}, \& {Gerhard}}]{Perez-Villegas2017}
	{P{\'e}rez-Villegas}, A., {Portail}, M., {Wegg}, C., \& {Gerhard}, O. 2017,
	\apjl, 840, L2
	
	\bibitem[{{Pichardo} {et~al.}(2004){Pichardo}, {Martos}, \&
		{Moreno}}]{Pichardo2004}
	{Pichardo}, B., {Martos}, M., \& {Moreno}, E. 2004, \apj, 609, 144
	
	\bibitem[{{Portail} {et~al.}(2015){Portail}, {Wegg}, {Gerhard}, \&
		{Martinez-Valpuesta}}]{Portail2015}
	{Portail}, M., {Wegg}, C., {Gerhard}, O., \& {Martinez-Valpuesta}, I. 2015,
	\mnras, 448, 713
	
	\bibitem[{{Robin} {et~al.}(2017){Robin}, {Bienaym{\'e}},
		{Fern{\'a}ndez-Trincado}, \& {Reyl{\'e}}}]{Robin2017}
	{Robin}, A.~C., {Bienaym{\'e}}, O., {Fern{\'a}ndez-Trincado}, J.~G., \&
	{Reyl{\'e}}, C. 2017, ArXiv e-prints
	
	\bibitem[{{Robin} {et~al.}(2012){Robin}, {Marshall}, {Schultheis}, \&
		{Reyl{\'e}}}]{Robin2012}
	{Robin}, A.~C., {Marshall}, D.~J., {Schultheis}, M., \& {Reyl{\'e}}, C. 2012,
	\aap, 538, A106
	
	\bibitem[{{Robin} {et~al.}(2003){Robin}, {Reyl{\'e}}, {Derri{\`e}re}, \&
		{Picaud}}]{Robin2003}
	{Robin}, A.~C., {Reyl{\'e}}, C., {Derri{\`e}re}, S., \& {Picaud}, S. 2003,
	\aap, 409, 523
	
	\bibitem[{{Robin} {et~al.}(2014){Robin}, {Reyl{\'e}}, {Fliri}, {Czekaj},
		{Robert}, \& {Martins}}]{Robin2014}
	{Robin}, A.~C., {Reyl{\'e}}, C., {Fliri}, J., {et~al.} 2014, \aap, 569, A13
	
	\bibitem[{{Romero-G{\'o}mez} {et~al.}(2015){Romero-G{\'o}mez}, {Figueras},
		{Antoja}, {Abedi}, \& {Aguilar}}]{Romero-Gomez2015}
	{Romero-G{\'o}mez}, M., {Figueras}, F., {Antoja}, T., {Abedi}, H., \&
	{Aguilar}, L. 2015, \mnras, 447, 218
	
	\bibitem[{{Simion} {et~al.}(2017){Simion}, {Belokurov}, {Irwin}, {Koposov},
		{Gonzalez-Fernandez}, {Robin}, {Shen}, \& {Li}}]{Simion2017}
	{Simion}, I.~T., {Belokurov}, V., {Irwin}, M., {et~al.} 2017, ArXiv e-prints
	
	\bibitem[{{Sofue}(2015)}]{Sofue2015}
	{Sofue}, Y. 2015, \pasj, 67, 75
	
	\bibitem[{{Tang} {et~al.}(2017){Tang}, {Fern\'andez-Trincado}, {Geisler},
		{Zamora}, {M\'esz\'aros}, \& {Masseron}}]{Tang2017}
	{Tang}, B., {Fern\'andez-Trincado}, J.~G., {Geisler}, D., {et~al.} 2017,
	submitted, 1, 1
	
	\bibitem[{{Wegg} \& {Gerhard}(2013)}]{Wegg2013}
	{Wegg}, C. \& {Gerhard}, O. 2013, \mnras, 435, 1874
	
	\bibitem[{{Zasowski} {et~al.}(2016){Zasowski}, {Ness}, {Garc{\'{\i}}a
			P{\'e}rez}, {Martinez-Valpuesta}, {Johnson}, \& {Majewski}}]{Zasowski2016}
	{Zasowski}, G., {Ness}, M.~K., {Garc{\'{\i}}a P{\'e}rez}, A.~E., {et~al.} 2016,
	\apj, 832, 132
	
\end{thebibliography}
\end{document}